\documentclass[a4paper,fleqn]{cas-sc}

\usepackage[numbers,sort&compress]{natbib}
\usepackage{graphicx}
\usepackage{amsmath,amssymb}
\usepackage{booktabs}
\usepackage{textcomp}
\usepackage{xspace}
\usepackage[section]{placeins}

\usepackage{xpatch}
\AtBeginDocument{}

\begin{document}
\let\WriteBookmarks\relax
% Allow text and floats to mix on the same page
\renewcommand{\topfraction}{0.9}
\renewcommand{\bottomfraction}{0.9}
\renewcommand{\textfraction}{0.05}
\renewcommand{\floatpagefraction}{0.7}

% Short title
\shorttitle{A multi-stage soft computing framework for liver cirrhosis}

% Short authors
\shortauthors{X.~Huang et~al.}

% Main title
\title[mode=title]{A multi-stage soft computing framework for complex disease modelling and decision support: A liver cirrhosis case study}

%================== Authors ==================
\author[1]{Xueyuan Huang}
\fnmark[1]
\credit{Writing -- review \& editing, Writing -- original draft, Supervision, Conceptualization}
\affiliation[1]{organization={Department of Hepatobiliary Surgery, the Second Affiliated Hospital of Chongqing Medical University},
                city={Chongqing},
                postcode={400010},
                country={China}}

\author[2]{Yuheng Wang}
\fnmark[1]
\credit{Writing -- original draft, Software, Methodology, Investigation, Conceptualization}
\affiliation[2]{organization={State Key Laboratory of Respiratory Health and Multimorbidity, Institute of Basic Medical Sciences \& School of Basic Medicine, Chinese Academy of Medical Sciences \& Peking Union Medical College},
                city={Beijing},
                postcode={100005},
                country={China}}

\author[3]{Yuanzhi He}
\fnmark[1]
\credit{Writing -- original draft, Visualization, Software, Formal analysis}
\affiliation[3]{organization={School of Computer Science and Informatics, Cardiff University},
                city={Wales},
                postcode={CF24 0DE},
                country={United Kingdom}}

\author[4]{Siqi Gou}
\credit{Writing -- review \& editing, Validation, Supervision, Formal analysis}
\affiliation[4]{organization={Department of Medical Oncology, the First Hospital of China Medical University},
                city={Shenyang},
                postcode={110001},
                country={China}}

\author[2]{Lu Bai}
\credit{Visualization, Validation}

\author[2]{Wenqian Wu}
\credit{Visualization, Validation}

\author[5]{Peifeng Liu}
\credit{Writing -- review \& editing}
\affiliation[5]{organization={Institute for Innovation and Development, Tsinghua University},
                city={Beijing},
                postcode={100086},
                country={China}}

\author[2]{Aijia Wang}
\credit{Data curation}

\author[2]{Tianhui Fan}
\credit{Resources, Project administration}

\author[6]{Ze Zhou}
\affiliation[6]{organization={Department of Liver Surgery, Peking Union Medical College Hospital, Chinese Academy of Medical Sciences \& Peking Union Medical College},
                city={Beijing},
                country={China}}

\author[3]{Jiayu Xu}
\cormark[1]
\ead{XuJ79@cardiff.ac.uk}
\credit{Visualization, Validation, Resources, Project administration, Data curation, Writing -- review \& editing}

\cortext[1]{Corresponding author.}
\fntext[1]{These authors contributed equally to this work.}

%================== Abstract ==================
\begin{abstract}
Liver cirrhosis is a major public health problem worldwide. It causes millions of deaths each year in both developed and developing countries. Timely detection and aggressive treatment can significantly improve patients' quality of life. Modelling complex diseases using biomedical data remains a challenging computational problem due to high dimensionality, strong feature correlations, noise, and limited labelled samples. Conventional Machine Learning (ML) pipelines often struggle with robustness, interpretability, and generalisation under such conditions. In this study, we propose a ML--driven multi-stage decision framework for complex disease modelling and therapeutic exploration. The framework integrates single-cell transcriptomic profiling, high-dimensional network-based feature stabilisation, multi-model learning, deep representation construction, and post-hoc decision support. Specifically, single-cell sequencing data were analysed to identify key cellular subpopulations, followed by high-dimensional weighted gene co-expression network analysis (hdWGCNA) to stabilise gene modules under sparsity and noise. To enhance non-linear feature interaction modelling, tabular molecular features were restructured into two-dimensional disease maps and analysed using a CNN. Finally, molecular docking was incorporated as a decision-support module to evaluate candidate therapeutic compounds. Using liver cirrhosis as a representative case study, the proposed framework identified a disease-associated endothelial cell subpopulation and extracted seven robust signature genes (HSPB1, GADD45A, CLDN5, ATP1B3, C1QBP, ENPP2, and PARL). The CNN-based representation learning module demonstrated superior classification performance compared with conventional pipelines. The proposed framework is disease-agnostic and can be readily extended to other omics-driven biomedical applications involving uncertainty, heterogeneity, and limited samples.
\end{abstract}

%================== Keywords ==================

\maketitle

%================== Introduction ==================
\section{Introduction}\label{sec:intro}

Liver cirrhosis remains a major global health challenge, with substantial morbidity and mortality across both developed and developing countries~\cite{ref1,ref2}. Despite continuous advances in clinical management, early diagnosis and effective therapeutic intervention are often hindered by heterogeneous aetiology, complex disease progression, and limited availability of robust predictive tools~\cite{ref3}. From a computational perspective, cirrhosis-related biomedical data are typically characterised by high dimensionality, strong feature correlations, biological noise, and limited labelled samples~\cite{ref4}. These properties pose significant challenges to conventional statistical and ML approaches, which may suffer from unstable feature selection and limited generalisation under such conditions. Consequently, there is a growing need for computational frameworks that can robustly model disease complexity and uncertainty to support early diagnosis and data-driven therapeutic decision-making.

At the cellular level, liver cirrhosis is driven by coordinated interactions among multiple cell types within the hepatic microenvironment. Hepatic stellate cells have been extensively studied due to their central role in extracellular matrix deposition and fibrotic remodelling~\cite{ref5}. However, increasing evidence indicates that endothelial cells play a critical regulatory role in vascular remodelling, inflammatory signalling, and fibrotic niche formation during chronic liver injury~\cite{ref6}. Dysfunctional endothelial activity contributes to abnormal angiogenesis, portal hypertension, and sustained activation of pro-fibrotic pathways~\cite{ref7}. Moreover, endothelial cells interact dynamically with hepatic stellate cells through paracrine signalling, amplifying fibrogenesis and disease progression~\cite{ref8}. The regulatory position of endothelial cells at the interface between vascular, immune, and stromal compartments makes them a particularly informative target for data-driven modelling of cirrhosis-associated molecular patterns~\cite{ref9}.

Recent advances in high-throughput sequencing technologies, particularly single-cell RNA sequencing, provide unprecedented opportunities to characterise disease-associated cellular heterogeneity at high resolution~\cite{ref10}. Nevertheless, the resulting data are inherently high-dimensional and sparse, introducing additional computational challenges for downstream analysis and predictive modelling~\cite{ref11}. Effective utilisation of single-cell data therefore requires feature stabilisation strategies that can mitigate noise while preserving biologically meaningful structure~\cite{ref12}.

In this study, liver cirrhosis is adopted as a representative case to develop and evaluate a machine learning--driven analytical framework for complex disease modelling. Single-cell transcriptomic analysis is first employed to identify disease-associated endothelial subpopulations, reducing biological confounding at the cellular level~\cite{ref13}. High-dimensional weighted gene co-expression network analysis (hdWGCNA) is then applied to stabilise gene modules under sparsity and noise~\cite{ref14}. Representative characterisation genes are selected using sparse learning via the least absolute shrinkage and selection operator (LASSO)~\cite{ref15} and subsequently evaluated across multiple ML models to ensure robustness against model-specific inductive bias. To further capture non-linear interactions beyond conventional tabular modelling, molecular signatures and immune infiltration features are restructured into two-dimensional representations and analysed using a CNN--based classification module~\cite{ref16}. Molecular docking is incorporated as a downstream decision-support step to prioritise candidate therapeutic compounds.

Together, this unified framework integrates single-cell analysis, network-based feature stabilisation, sparse learning, representation-enhanced deep learning, and post-hoc decision support to address the computational challenges of modelling liver cirrhosis under uncertainty~\cite{ref17,ref18}. While validated using cirrhosis data, the proposed framework is disease-agnostic and can be readily extended to other complex diseases characterised by high-dimensional, heterogeneous biomedical data.

%================== Related Work ==================
\section{Related Work}\label{sec:related}

ML has been increasingly explored to support liver cirrhosis diagnosis, severity assessment, and prognosis prediction using collected clinical variables. While broader hepatology reviews highlight the expanding use of ML across chronic liver disease management, they also emphasise common barriers such as dataset heterogeneity, limited external validation, and interpretability requirements for clinical adoption~\cite{ref19,ref20}. The robust deployment requires careful handling of distribution shift and evaluation design beyond single-cohort performance reporting~\cite{ref21}.

CNN-based models have been investigated in liver imaging applications, benefiting from naturally grid-structured inputs. For instance, BCNN has been reported to achieve strong performance in differentiating liver lesions on dynamic contrast-enhanced CT, illustrating the effectiveness of CNNs in hepatobiliary imaging tasks~\cite{ref14}. Beyond imaging, deep learning has also been explored for cirrhosis detection using other structured biomedical signals, such as ECG-based modelling for cirrhosis-related signatures~\cite{ref22}. These works collectively motivate representation-aware modelling when raw biomedical inputs are not inherently image-like.

In transcriptomics-driven cirrhosis research, feature selection and classical classifiers remain common choices in high-dimensional, low-sample settings. To improve robustness and interpretability, network-based approaches such as WGCNA are frequently used to identify co-expression modules and prioritise hub genes as stable candidate signatures~\cite{ref23}. In parallel, immune microenvironment estimation from bulk expression profiles has become an important component of liver disease modelling; CIBERSORT is a widely used deconvolution framework for this purpose. Systematic benchmarking further shows that deconvolution accuracy and bias depend on the chosen method and data characteristics, supporting careful feature integration under noisy conditions~\cite{ref24}.

Beyond conventional ML pipelines, soft computing frameworks have been widely explored for disease modelling due to their ability to tolerate uncertainty, imprecision, and incomplete information. Fuzzy logic--driven models represent one of the earliest and most extensively studied soft computing approaches in biomedical applications~\cite{ref25}. By encoding expert knowledge into linguistic rules and membership functions, fuzzy classifiers have been applied to disease diagnosis, risk stratification, and clinical decision support~\cite{ref26}, particularly in scenarios where crisp decision boundaries are difficult to define. However, such approaches often rely on handcrafted rule bases or domain-specific tuning, which may limit scalability and generalisability when applied to high-dimensional omics data.

Evolutionary computation has also been frequently adopted within soft computing frameworks to optimise feature selection and model parameters for disease classification~\cite{ref27}. Genetic algorithms and related evolutionary strategies have been used to search high-dimensional feature spaces, identify informative biomarker subsets, and improve classification performance under limited sample conditions. While these methods are effective in global optimisation and feature subset discovery, they are typically employed as standalone optimisation components rather than as part of an integrated, end-to-end disease modelling framework. As a result, feature stabilisation, model learning, and downstream interpretation are often treated as loosely connected steps.

More recently, hybrid soft computing frameworks have emerged that combine fuzzy systems, neural networks, and evolutionary algorithms to address complex biomedical problems~\cite{ref28}. These hybrid models demonstrate improved flexibility compared with single-paradigm approaches, yet they frequently focus on optimising a specific predictive component rather than structuring the entire modelling pipeline across multiple abstraction levels. In addition, many existing soft computing frameworks are designed for specific diseases or data modalities, limiting their adaptability to heterogeneous biomedical contexts.

In contrast to existing disease modelling approaches, the present study proposes a unified multi-stage soft computing framework that integrates biological data abstraction, network-based feature stabilisation, sparse learning, representation-enhanced deep learning, and post-hoc decision support into a coherent computational pipeline. While prior cirrhosis-related ML studies have established effective building blocks across clinical prediction, DL in imaging or signal domains, and transcriptomic biomarker discovery, these components are often applied as loosely connected steps, which can limit robustness and methodological clarity under uncertainty. Rather than relying on expert-defined rules or single-stage optimisation, the proposed framework organises heterogeneous learning strategies across successive stages, with each stage addressing a distinct computational challenge. Importantly, the framework is disease-agnostic by design: individual modules can be independently adapted or replaced according to data modality and application context, enabling flexible deployment across different complex diseases. This modular, multi-stage integration distinguishes the proposed framework from existing fuzzy logic--driven or evolutionarily optimised disease modelling approaches and aligns with the core principles of soft computing for decision-making under uncertainty.

%================== Materials and Methods ==================
\section{Materials and Methods}\label{sec:methods}

\subsection{Data Sources and Experimental Design}
Single-cell RNA sequencing data for liver cirrhosis were obtained from the Gene Expression Omnibus (GEO) database under accession number GSE136103. These data were used to support cellular-level feature construction within the proposed framework. To enable independent model training and evaluation, bulk transcriptomic datasets were retrieved from GEO. GSE14323 was used as the training cohort, while GSE6764 served as an independent testing cohort.

\subsection{Single-Cell Data Processing and Disease-Relevant Subpopulation Identification}
Single-cell transcriptomic data were processed using the Seurat framework. Quality control, normalisation, and identification of highly variable genes were performed following standard procedures. Batch effects across samples were corrected using the Harmony algorithm.

Cell type annotation was conducted using reference datasets from the celldex resource. Endothelial cell populations were selected as disease-relevant subpopulations and used as inputs for downstream network-based feature stabilisation. This step aims to reduce biological confounding and constrain feature extraction to informative cellular states.

\subsection{Network-Based Feature Stabilisation}
To stabilise high-dimensional and sparse gene expression profiles, high-dimensional weighted gene co-expression network analysis (hdWGCNA) was applied to endothelial cell expression data. Genes expressed in at least 5\% of cells were retained. Metacells were constructed prior to network construction to reduce sparsity. A soft-thresholding power was selected to approximate scale-free topology. Gene modules were identified via hierarchical clustering, and hub genes were ranked based on module eigengene-based connectivity (kME). Genes from disease-associated modules were retained as candidate molecular features.

\subsection{Sparse Feature Selection and Classical Machine Learning}
Candidate features derived from network-based stabilisation were subjected to sparse feature selection using univariate logistic regression followed by least absolute shrinkage and selection operator (LASSO) regression. Selected features were standardised prior to modelling. Seven classical machine learning classifiers---k-nearest neighbours, linear discriminant analysis, logistic regression, na\"{\i}ve Bayes, random forest, decision tree, and support vector machine---were implemented for comparative evaluation. Model training was conducted using repeated stratified five-fold cross-validation on the training cohort, and final performance was assessed on the independent testing cohort.

\subsection{Representation-Enhanced Deep Learning}
To capture non-linear interactions beyond conventional tabular modelling, structured two-dimensional representations were constructed using selected molecular features and immune infiltration scores. These representations were designed to provide CNN-compatible inputs while constraining model capacity. A CNN is applied to the constructed representations for classification. The CNN operates on structured feature maps rather than raw high-dimensional data, enabling representation learning under limited sample conditions.

\subsection{Overall Multi-stage Computational Framework}
The proposed approach is formulated as a multi-stage computational framework that progressively transforms high-dimensional and heterogeneous biomedical data into structured representations suitable for predictive modelling and decision support. From a computational perspective, the framework follows a hierarchical abstraction strategy, in which raw data are incrementally converted into stabilised features, compact representations, and final decision-level outputs.

At the feature abstraction level, cellular-scale filtering and network-based stabilisation act as dimensionality reduction and noise-tolerant preprocessing mechanisms, constraining downstream learning to coherent and reproducible feature subspaces. At the modelling level, sparse learning and classical machine learning algorithms operate on compact feature sets to establish baseline decision boundaries under limited-sample conditions. These models provide complementary inductive biases and serve as reference learners for subsequent representation-enhanced modelling.

At the representation learning level, structured two-dimensional disease maps enable CNN to model higher-order and non-linear feature interactions that are difficult to capture using conventional tabular approaches. Importantly, the CNN module does not directly process raw high-dimensional data but operates on constrained representations derived from prior stabilisation and selection stages, thereby limiting effective model capacity and reducing overfitting risk.

%================== Results ==================
\section{Results}\label{sec:results}

\subsection{Flowchart and mechanism of endothelial cell-derived liver cirrhosis}
\begin{figure}[!htbp]
  \centering
  \includegraphics[width=\linewidth,height=0.78\textheight,keepaspectratio]{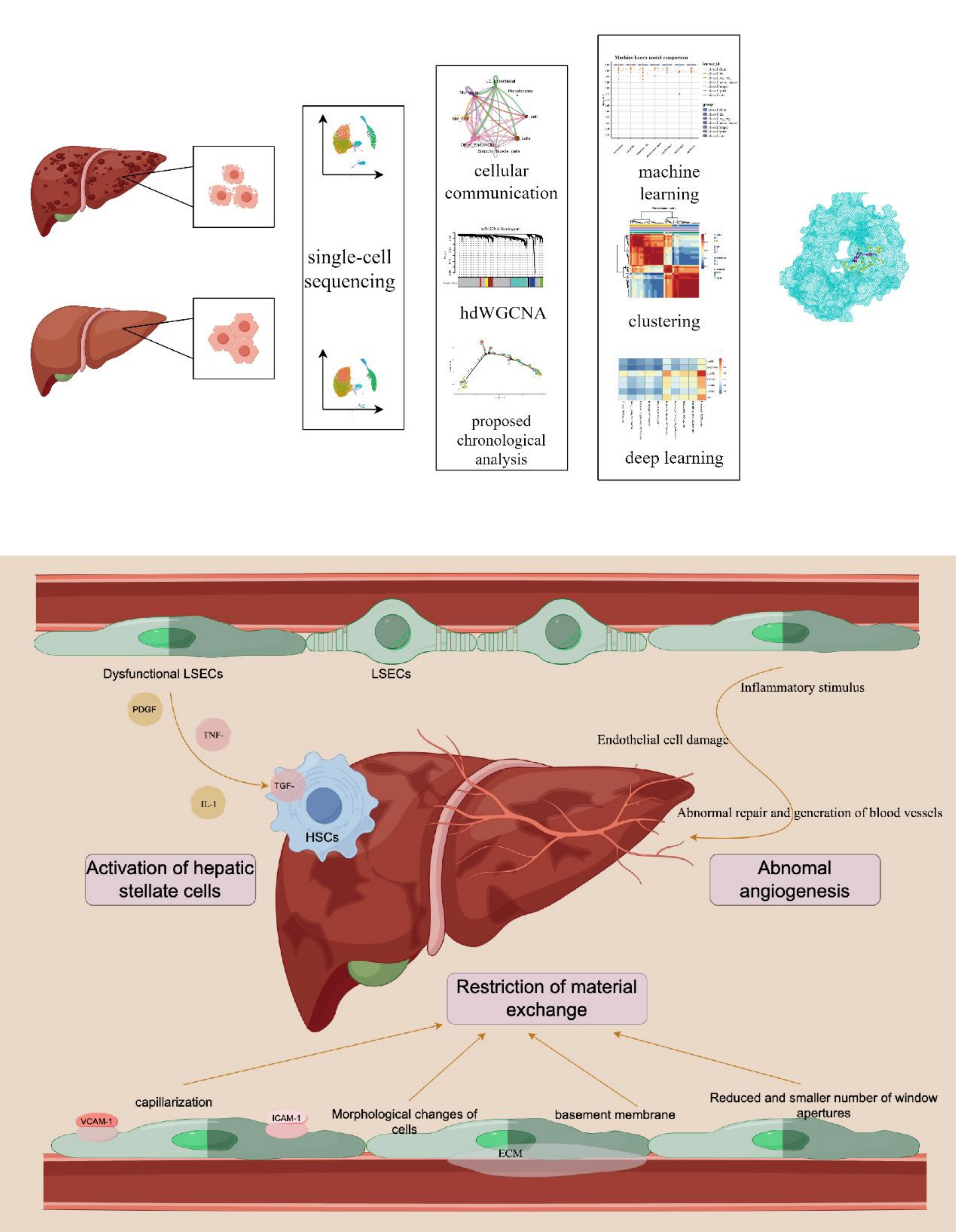}
  \caption{Flow chart of the study and the mechanism drawing of endothelial cells causing liver cirrhosis.}
  \label{fig:fig1}
\end{figure}

A flowchart was used to show the whole process of this research. In addition, we mapped the mechanisms of endothelial cells in liver cirrhosis (Figure~\ref{fig:fig1}).

\subsection{Cell type annotation and key subpopulations in liver cirrhosis single-cell sequencing data}
Single-cell RNA sequencing data from liver cirrhotic tissue and normal liver tissue were obtained from the GEO database. After data loading, a Seurat object containing 15{,}238 cells was constructed. Following rigorous quality control based on predefined filtering criteria, 14{,}469 high-quality cells were retained for subsequent analysis. These preprocessing steps ensured the reliability of downstream clustering and subpopulation identification.

Dimensionality reduction and unsupervised clustering were then performed to explore cellular heterogeneity. Principal component analysis (PCA) was applied to project high-dimensional gene expression profiles into a low-dimensional space for efficient visualisation and clustering. To minimise technical variability arising from batch effects, the Harmony algorithm was employed to correct inter-sample differences caused by processing and measurement conditions. Based on variance explained and clustering stability, the first five principal components were selected for downstream analysis.

\begin{figure}[!htbp]
  \centering
  \includegraphics[width=\linewidth,height=0.78\textheight,keepaspectratio]{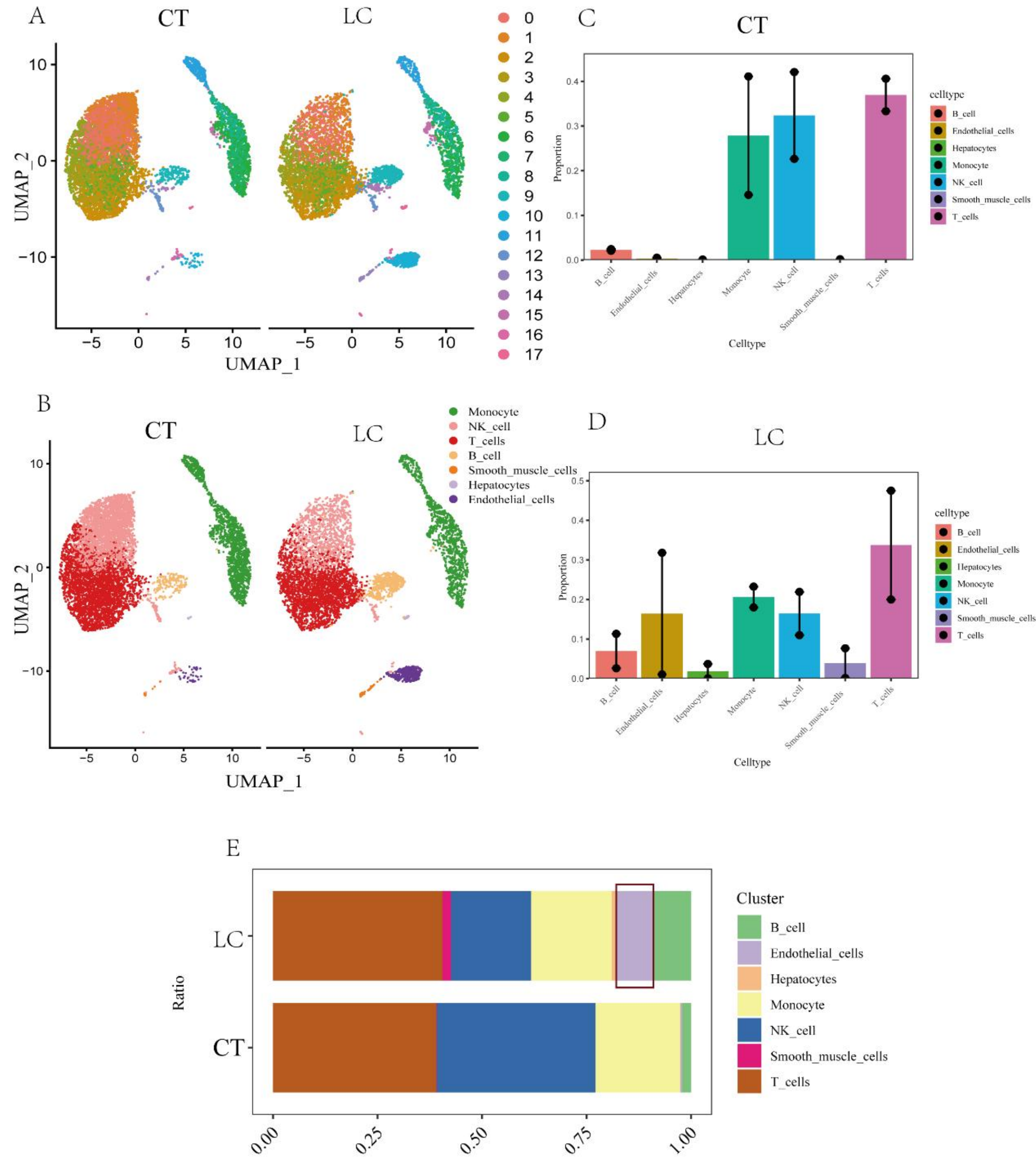}
  \caption{Single-cell sequencing identifies key cell subpopulations. (A) Single-cell clustering results for normal and cirrhotic samples. (B) Single-cell annotation results for normal and cirrhotic samples. Different colours represent different cell subpopulations. (C, D) Bar graphs show the number of cells in each subpopulation: (C) control group; (D) liver cirrhosis group. (E) Horizontal bar graphs were used to demonstrate differences in cell subpopulations between the normal and disease groups.}
  \label{fig:fig2}
\end{figure}

Using these corrected low-dimensional representations, 17 distinct cell clusters were identified and visualised. Cell type annotation was performed using reference datasets provided by the celldex package, enabling accurate mapping of clusters to known cell identities (Figure~\ref{fig:fig2}A). Through this data-driven annotation process, seven major cell types were consistently identified across cirrhotic and control samples (Figure~\ref{fig:fig2}B).

To facilitate a more intuitive comparison of cellular composition between disease and control groups, bar plots were generated to illustrate the relative abundance of each annotated cell type (Figure~\ref{fig:fig2}C--E). Pronounced differences in cell population distributions were observed between the two groups, reflecting disease-associated cellular remodelling. Notably, endothelial cells exhibited a striking shift in abundance: they were nearly absent in control samples but markedly enriched in cirrhotic tissues. This pronounced and consistent alteration suggests that endothelial cells represent a key disease-associated subpopulation.

From a computational perspective, the emergence of endothelial cells as a discriminative subpopulation highlights their potential as stable and informative features for downstream modelling. Therefore, endothelial cells were selected for subsequent in-depth analysis, serving as a critical entry point for network-based feature extraction, ML classification, and deep representation learning in the proposed ML framework.

\subsection{Key subgroups in endothelial cells}
\begin{figure}[!htbp]
  \centering
  \includegraphics[width=\linewidth,height=0.78\textheight,keepaspectratio]{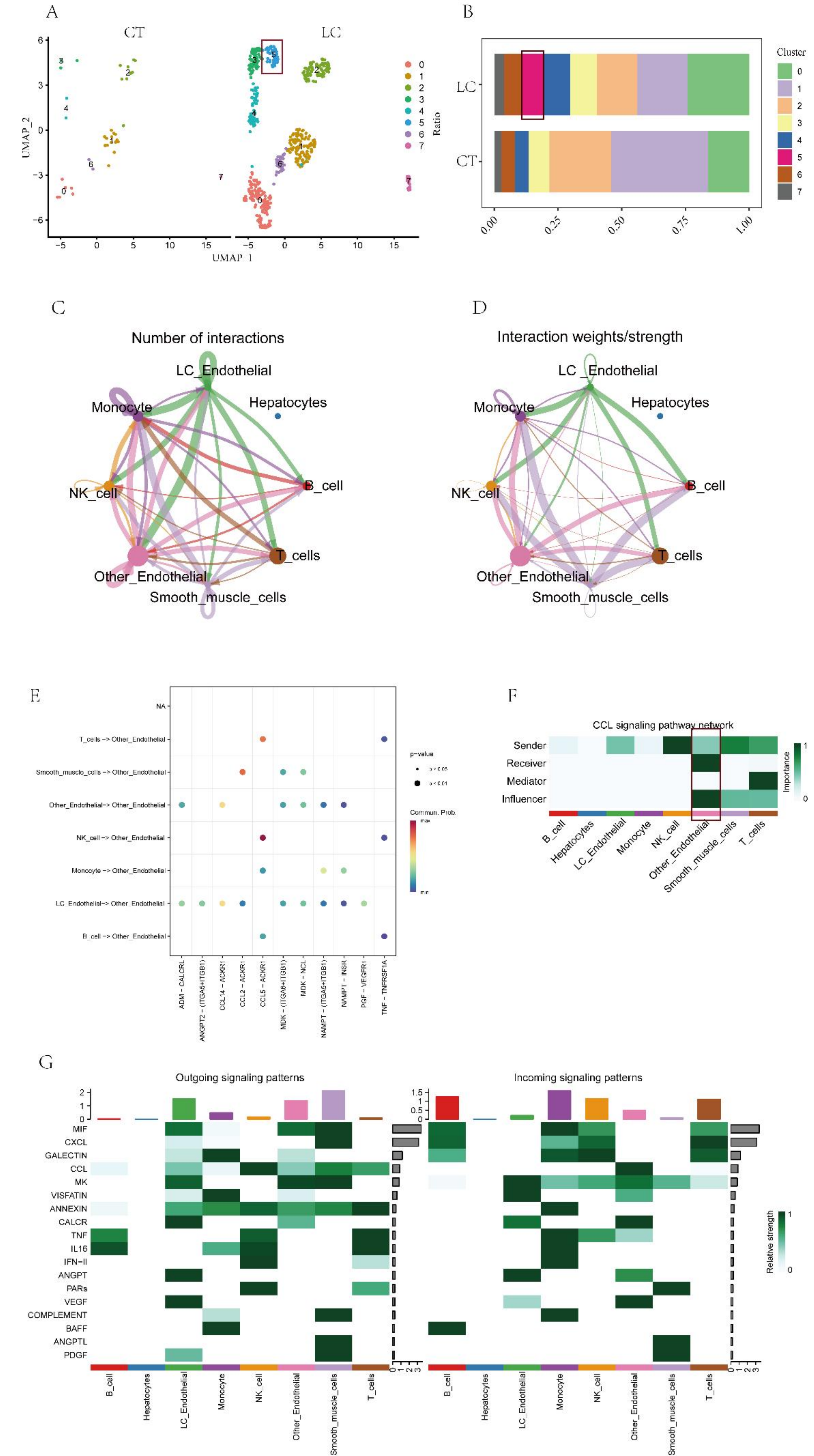}
  \caption{Identification of key subpopulations in endothelial cells and cellular communications. (A) Cell clustering in endothelial cells. The picture on the left shows the normal group; the disease group is shown on the right. (B) Horizontal bar chart shows subgroup differences between normal and disease groups. (C) The spherical linear plot demonstrates the amount of cellular communication between different cell communities; thicker line segments represent a greater number of cellular communications. (D) Weighting of cellular communication between different cellular communities; the thickness of the line segment represents the strength of the communication. (E) Pathways in other-Endothelial cell communication. (F) Role of other-Endothelial cell communities in the CCL signalling pathway. (G) Cellular communication in different cell communities.}
  \label{fig:fig3}
\end{figure}

Following the identification of endothelial cells as a disease-associated population, we further investigated their internal heterogeneity to determine whether distinct endothelial subgroups were involved in cirrhosis progression. To this end, endothelial cells were extracted and subjected to independent dimensionality reduction and unsupervised clustering analysis (Figure~\ref{fig:fig3}A). This secondary clustering strategy enabled a more fine-grained characterisation of endothelial cell states under disease conditions.

Notably, a distinct endothelial cell cluster emerged predominantly in cirrhotic samples, while being absent or minimally represented in control tissues (Figure~\ref{fig:fig3}B). Based on this disease-specific enrichment pattern, this cluster was defined as LC-Endothelial, whereas the remaining endothelial cells were collectively referred to as other-Endothelial. This data-driven subdivision suggests that endothelial cells in cirrhosis are not a homogeneous population, but instead comprise functionally divergent subgroups associated with disease status.

To further explore the functional relevance of these endothelial subgroups within the multicellular disease microenvironment, intercellular communication analysis was performed using the CellChat framework. Interaction networks among cell populations were inferred and visualised to assess both the number and strength of cell--cell communications. Circular network diagrams generated using the \texttt{netVisual\_circle} function illustrated the interaction structure across cell populations. LC-Endothelial cells exhibited a high number of interactions with multiple cell types, indicating their central involvement in the communication network (Figure~\ref{fig:fig3}C). LC-Endothelial cells also displayed strong interaction weights, reinforcing their dominant position within the inferred signalling architecture (Figure~\ref{fig:fig3}D).

From a network perspective, these results indicate that LC-Endothelial cells function as key communication hubs in cirrhotic tissue. Their extensive connectivity suggests that they may act as relay nodes coordinating information flow between endothelial cells and other disease-relevant cell populations, thereby contributing to global network reorganisation during cirrhosis progression.

In addition to LC-Endothelial cells, other-Endothelial cells also demonstrated notable pathway-specific activity. In particular, other-Endothelial cells showed prominent involvement in the C--C motif chemokine ligand (CCL) signalling pathway (Figure~\ref{fig:fig3}E). To further quantify the functional roles of cell populations within this pathway, centrality and role analyses were performed. Other-Endothelial cells exhibited high centrality scores and were assigned multiple roles, including Sender, Receiver, and Influencer, within the CCL signalling network (Figure~\ref{fig:fig3}F). These findings suggest that different endothelial subgroups contribute to cirrhosis through complementary communication patterns rather than redundant functions.

Finally, Figure~\ref{fig:fig3}G provides an integrated summary of the endothelial-centred communication landscape in liver cirrhosis. LC-Endothelial cells occupy a dominant position within the global interaction network, while other-Endothelial cells and additional cell populations play supporting but non-negligible roles. Together, these results highlight a structured and hierarchical organisation of endothelial subpopulations in cirrhosis, which forms the basis for subsequent network-based feature extraction and ML analysis. Detailed biological implications of these findings are further discussed in the Discussion section.

\subsection{HdWGCNA of endothelial cells}
To further characterise disease-associated transcriptional patterns within endothelial cells, high-dimensional weighted gene co-expression network analysis (hdWGCNA) was performed, with endothelial cells serving as the target population for network construction. This analysis aimed to stabilise high-dimensional single-cell gene expression data and identify coherent gene modules that capture coordinated biological signals under sparse and noisy conditions.

\begin{figure}[!htbp]
  \centering
  \includegraphics[width=\linewidth,height=0.78\textheight,keepaspectratio]{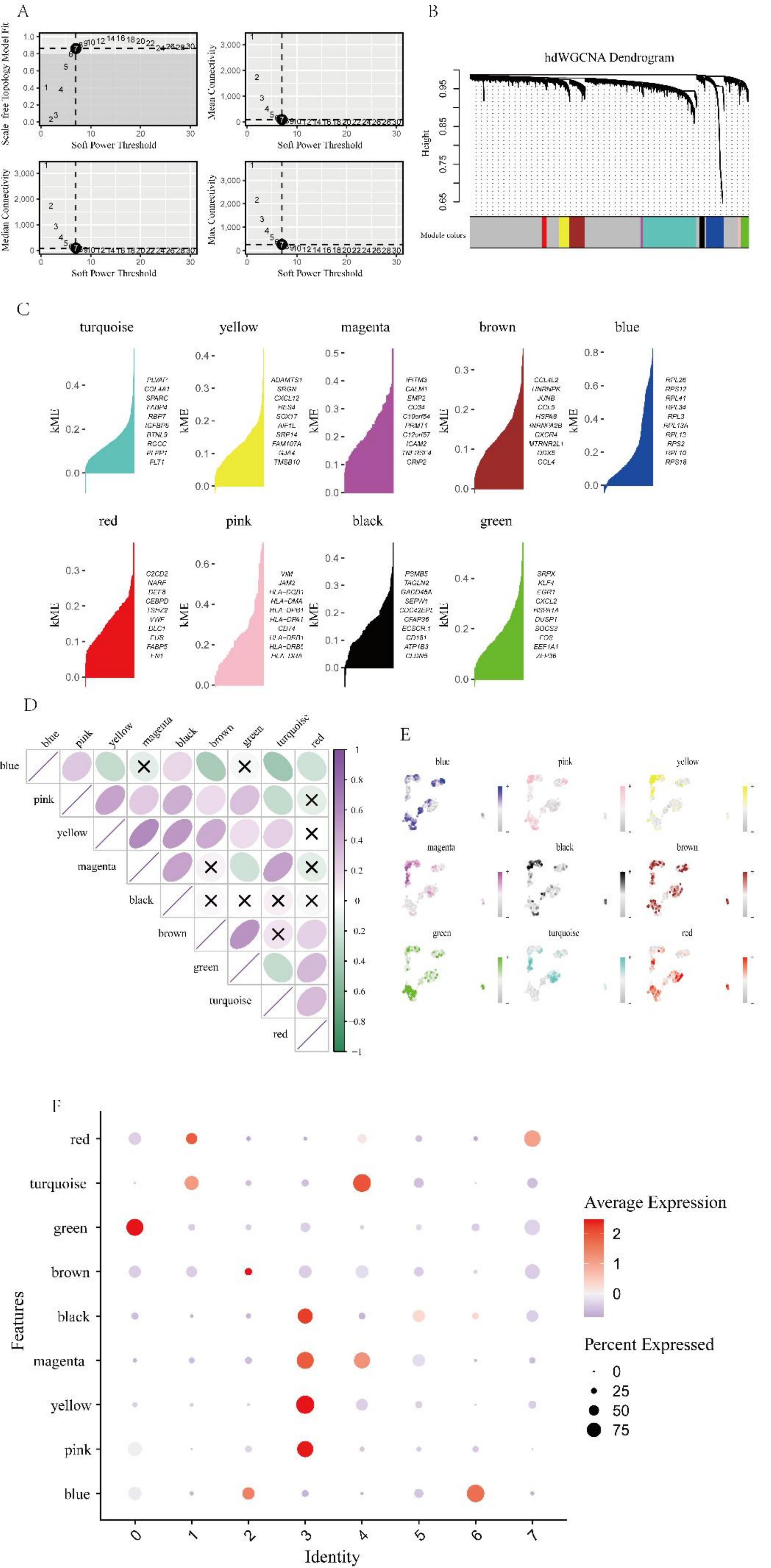}
  \caption{HdWGCNA of endothelial cells. (A) An appropriate soft-thresholding power for constructing a scale-free network. (B) Genes with similar functions are grouped into the same colour module. (C) Genes in each module were visualised and sorted by kME score. (D) Correlation between each module. (E) Expression positions of the genes in each module in the cirrhosis group. (F) Expression of seven cellular subpopulations of endothelial cells in the cirrhosis group by different colour modules.}
  \label{fig:fig4}
\end{figure}

An appropriate soft-thresholding power was determined using the \texttt{TestSoftPowers} function to ensure that the resulting co-expression network approximated scale-free topology (Figure~\ref{fig:fig4}A). Based on network fit and stability criteria, a soft-thresholding power of 7 was selected for subsequent network construction. Hierarchical clustering of genes based on topological overlap resulted in a dendrogram representation of the network structure, enabling visualisation of gene co-expression relationships (Figure~\ref{fig:fig4}B). Through this process, genes were grouped into nine distinct colour-coded modules, each representing a set of genes with highly correlated expression patterns.

To assess the internal structure of each module, genes were ranked according to their module eigengene-based connectivity (kME), which quantifies the correlation between individual gene expression profiles and the corresponding module eigengene. Higher kME values indicate greater relevance and centrality within a given module (Figure~\ref{fig:fig4}C). Module--module relationships were further evaluated by calculating pairwise correlations among module eigengenes, with statistically significant associations identified at a threshold of $P < 0.001$ (Figure~\ref{fig:fig4}D).

To link gene modules to specific endothelial subpopulations, expression scores were calculated using the UCell method for the top 25 hub genes within each module. Module feature scores, represented by hub module eigengenes (hMEs), were visualised using the \texttt{ModuleFeaturePlot} function. The resulting distributions illustrate the activity of each gene module across endothelial cells (Figure~\ref{fig:fig4}E). Notably, differential module activation patterns were observed among endothelial subgroups.

Comparative analysis across endothelial subpopulations revealed that the five-cell cluster defined as LC-Endothelial exhibited a distinct module activity profile (Figure~\ref{fig:fig4}F). In particular, genes belonging to the black module showed pronounced enrichment within the LC-Endothelial subpopulation. This selective activation indicates that the black module captures a disease-associated transcriptional program specific to LC-Endothelial cells.

Based on these network-driven results, hub genes from the LC-Endothelial--associated module were prioritised as candidate features for downstream modelling. These genes served as a stabilised and biologically coherent feature set for subsequent sparse selection and ML analyses.

\subsection{Proposed chronological analysis and enrichment analysis in endothelial cells}
\begin{figure}[!htbp]
  \centering
  \includegraphics[width=\linewidth,height=0.78\textheight,keepaspectratio]{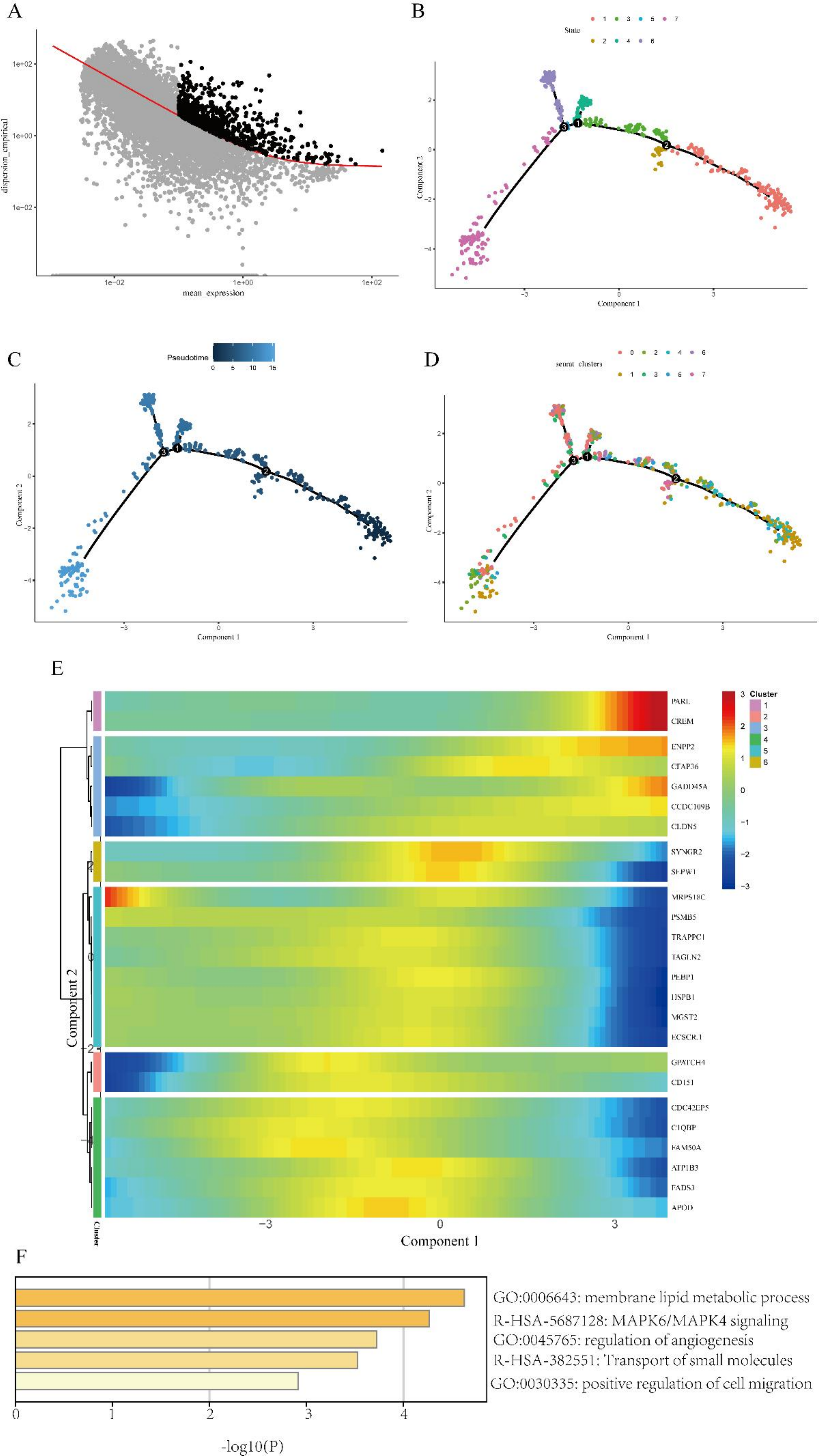}
  \caption{Proposed temporal analysis of endothelial cells and enrichment analysis. (A) Filter conditions: mean expression $\geq 0.1$, empirical dispersion $\geq 1$. Visualisation of the screened genes. (B,~C) Growth and developmental trajectories of endothelial cells. (D) Distributional position of seven subpopulations of endothelial cells in growth and developmental trajectories. (E) Changes in expression of LC-Endothelial subpopulation core genes in the proposed time series. (F) GO pathway enrichment analysis.}
  \label{fig:fig5}
\end{figure}

To further characterise the relative position of the LC-Endothelial subpopulation within the broader endothelial cell landscape, this analysis aimed to reconstruct pseudo-temporal trajectories and infer dynamic transcriptional changes associated with endothelial cell state transitions under cirrhotic conditions. Genes with low expression noise and high variability were selected for trajectory inference based on the following criteria: mean expression $\geq 0.1$ and empirical dispersion $\geq 1$ (Figure~\ref{fig:fig5}A). This filtering strategy ensured that informative and dynamically regulated genes were retained for downstream analysis.

Based on these selected genes, endothelial cell developmental trajectories were reconstructed. The resulting trajectory structure revealed a clear directional progression of endothelial cell states, characterised by a right-to-left transition pattern in low-dimensional space (Figure~\ref{fig:fig5}B,C). This trajectory provides a data-driven ordering of endothelial cells, enabling the positioning of subpopulations along a continuous pseudo-temporal axis rather than discrete clusters.

Subsequently, the seven identified endothelial subpopulations were mapped onto the inferred developmental trajectory. Notably, the five-cell LC-Endothelial subpopulation was predominantly localised to the middle and late stages of the trajectory (Figure~\ref{fig:fig5}D), suggesting its association with advanced or transitional endothelial states under disease conditions. This temporal positioning further supports the disease relevance of LC-Endothelial cells identified through clustering and network analyses.

To investigate dynamic gene regulation within LC-Endothelial cells, twenty-five core genes derived from the LC-Endothelial--associated module were selected for proposed chronological analysis. Pseudo-temporal heatmaps constructed for these genes demonstrated distinct and coordinated expression changes along the inferred trajectory (Figure~\ref{fig:fig5}E), indicating that these genes capture temporal transcriptional programs linked to endothelial cell state transitions.

In addition, functional enrichment analysis was performed on the twenty-five LC-Endothelial core genes using Gene Ontology (GO) annotations. The enrichment results highlighted biological processes related to membrane lipid metabolic processes (Figure~\ref{fig:fig5}F), which encompass intracellular chemical reactions involved in the synthesis, transformation, and catabolism of membrane lipids. These processes have been reported to be closely associated with cirrhosis-related cellular remodelling. Detailed biological implications of these enriched pathways are further discussed in the Discussion section.

\subsection{Machine learning model performance comparison}
\begin{figure}[!htbp]
  \centering
  \includegraphics[width=\linewidth,height=0.78\textheight,keepaspectratio]{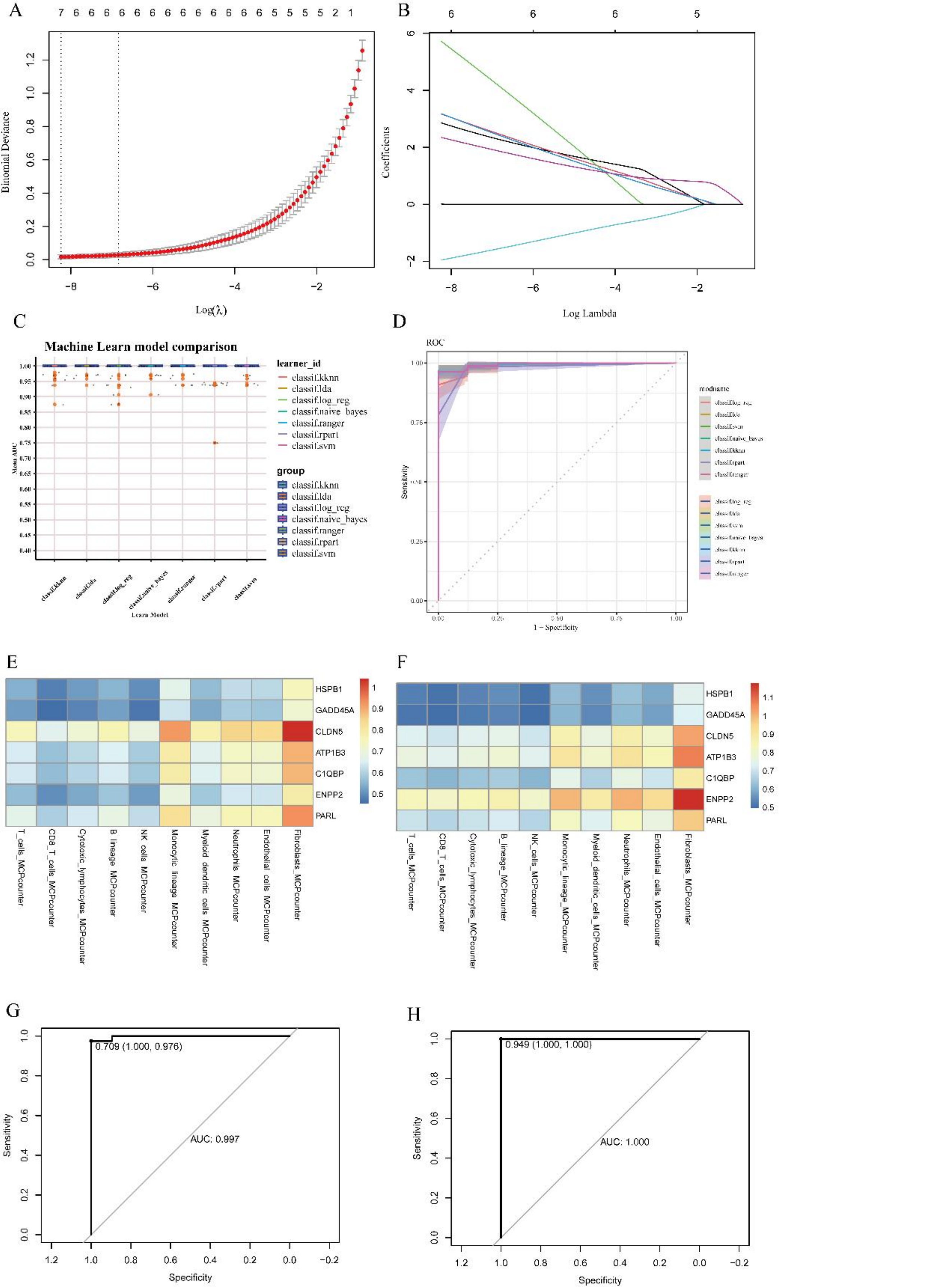}
  \caption{Machine learning and deep learning for the diagnosis of liver cirrhosis. (A) The LASSO Cox regression model shows partial likelihood deviance versus $\log(\lambda)$. (B) The lambda parameter indicates the coefficients of the extracted features; horizontal coordinates indicate the effect on the independent variable and vertical coordinates indicate the coefficients of the independent variable. (C) Comparison of seven machine learning models. (D) ROC graph for seven machine learning models. (E,~F) Deep learning 2D images of a patient in the training and test groups. (G,~H) Deep learning ROC curves for training and test groups.}
  \label{fig:fig6}
\end{figure}

To identify a compact and robust feature set from the LC-Endothelial gene space, least absolute shrinkage and selection operator (LASSO) regression was applied. The relationship between log-binomial deviance and $\log(\lambda)$ illustrates the model behaviour under varying penalisation strengths. Based on cross-validation, an optimal $\lambda$ value of 6 was selected. Under this constraint, seven representative feature genes were retained: HSPB1, GADD45A, CLDN5, ATP1B3, C1QBP, ENPP2, and PARL. These genes were considered stabilised molecular signatures associated with the LC-Endothelial subpopulation (Figure~\ref{fig:fig6}A,~B).

The selected seven genes were subsequently used as input features for seven classical ML classifiers, including k-nearest neighbours (kNN), linear discriminant analysis (LDA), logistic regression (LR), na\"{\i}ve Bayes (NB), random forest (RF), decision tree (DT), and support vector machine (SVM). Model performance was evaluated using receiver operating characteristic (ROC) curves and the area under the curve (AUC) metric. Training and validation were conducted on the GSE14323 dataset, while independent testing was performed on GSE6764, ensuring cross-platform and cross-cohort evaluation.

All seven classifiers demonstrated consistent discriminative performance based on the LC-Endothelial gene signatures (Figure~\ref{fig:fig6}C,~D), with detailed AUC values summarised in Table~\ref{tbl:tab1}. These results indicate that the selected gene set provides stable and transferable information for distinguishing cirrhotic from control samples across multiple learning paradigms, supporting its suitability for downstream modelling.

To further enhance diagnostic performance by capturing non-linear feature interactions, a CNN-based representation learning strategy was introduced. Immune and stromal cell infiltration profiles were estimated for both training and test datasets using MCPcounter deconvolution. Distinct variations in immune cell populations including T cells, CD8\textsuperscript{+} T cells, cytotoxic lymphocytes, B lineage cells, natural killer cells, and fibroblasts, were observed between cirrhotic and control samples (Figure~\ref{fig:fig6}E,F). Correlation analysis revealed that LC-Endothelial signature genes, such as HSPB1 and CLDN5, were associated with changes in fibroblast and endothelial cell abundance, suggesting coordinated gene--microenvironment relationships.

Based on these correlations, immune infiltration features and LC-Endothelial gene expression levels were integrated to construct two-dimensional ``disease maps'' for each sample. These representations were used as input to the CNN model, enabling the network to learn spatially structured feature interactions. The CNN was trained on the GSE58294 dataset and independently validated on GSE16561. As summarised in Table~\ref{tbl:tab1}, the CNN-based approach achieved higher classification performance than conventional ML models. Specifically, the model achieved an AUC of 0.997 in the training dataset and 1.000 in the independent test dataset, indicating strong discriminative capability. Sensitivity and specificity values for the training and test sets were 1.000 and 0.976, and 1.000 and 1.000, respectively (Figure~\ref{fig:fig6}G,H).

Overall, these results demonstrate that integrating sparse feature selection with representation-enhanced deep learning can substantially improve classification performance in complex disease settings. The CNN-based disease map approach provides a flexible framework for modelling non-linear interactions between molecular signatures and immune microenvironment features, offering a promising computational strategy for cirrhosis diagnosis and potentially other heterogeneous biomedical applications.

\begin{table}[!htb]
\caption{Machine learning and CNN algorithms for the diagnosis of cirrhotic liver patients. \texttt{auc\_train}: AUC on the training set. \texttt{auc\_test}: AUC on the test set. Sensitivity measures the ability of the model to correctly identify positive samples; specificity measures the ability of the model to correctly identify negative samples. Higher AUC, sensitivity and specificity values indicate stronger discrimination.}\label{tbl:tab1}
\centering
\begin{tabular*}{\tblwidth}{@{}LCCCC@{}}
\toprule
Learner & auc\_train & Sensitivity & Specificity & auc\_test \\
\midrule
Logistic Regression (LR)            & 1.0000 & 0.8767 & 0.9564 & 0.9853 \\
Linear Discriminant Analysis (LDA)  & 0.9989 & 0.9350 & 0.9753 & 0.9955 \\
Support Vector Machine (SVM)        & 1.0000 & 0.8933 & 0.9828 & 0.9943 \\
Naive Bayes (NB)                    & 0.9980 & 0.8933 & 0.9753 & 0.9949 \\
K-Nearest Neighbours (KNN)          & 0.9973 & 0.8933 & 0.9753 & 0.9923 \\
Decision Tree (DT)                  & 0.9884 & 0.9800 & 0.9761 & 0.9781 \\
Random Forest (RF)                  & 1.0000 & 0.8833 & 0.9953 & 0.9942 \\
Convolutional Neural Network (CNN)  & 0.9970 & 1.0000 & 0.9760 & 1.0000 \\
\bottomrule
\end{tabular*}
\end{table}

\subsection{Immune cell infiltration and gene expression patterns}
\begin{figure}[!htbp]
  \centering
  \includegraphics[width=\linewidth,height=0.78\textheight,keepaspectratio]{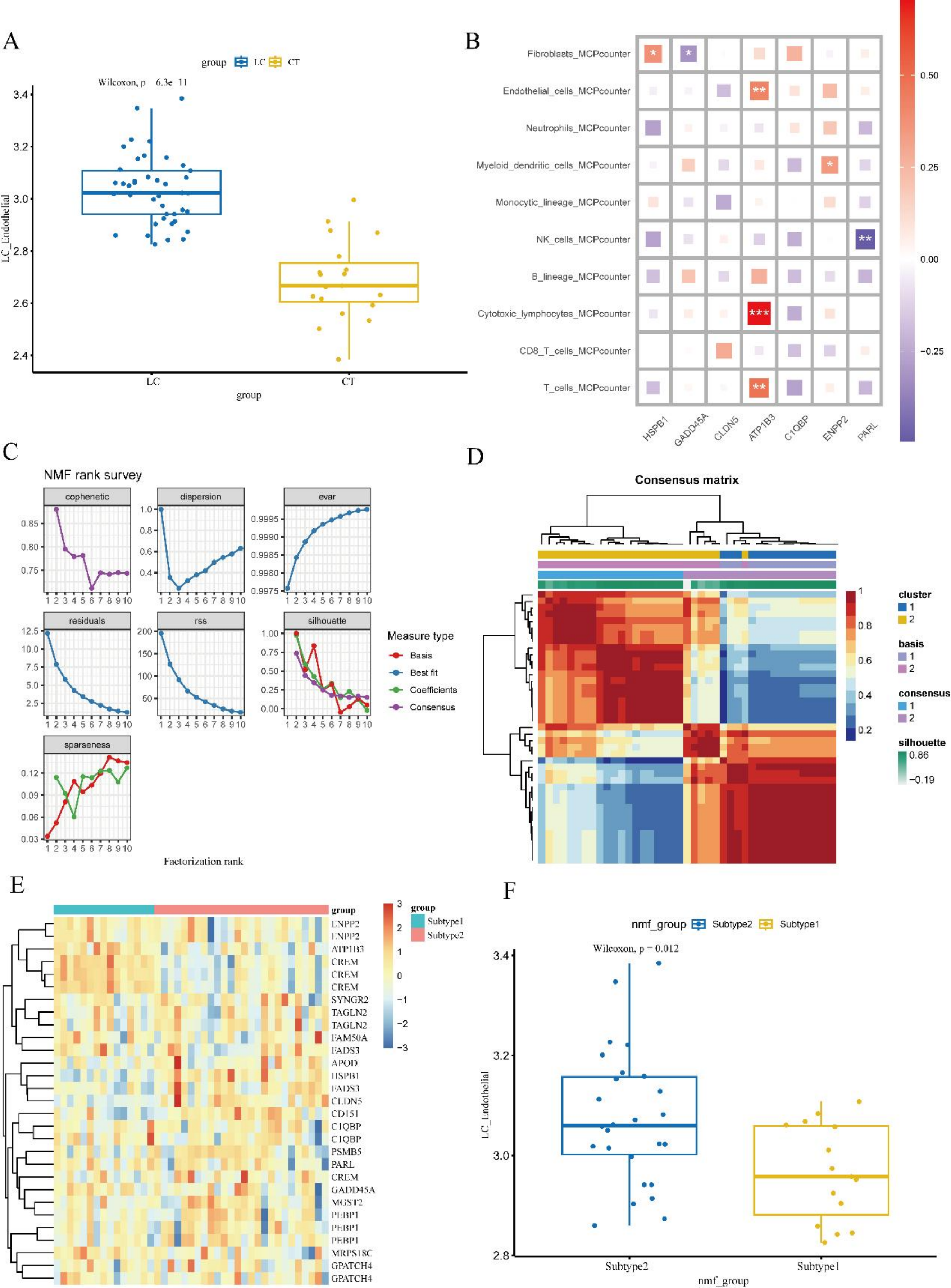}
  \caption{Cluster analysis. (A) Differences in the expression of 25 LC-endothelial genes in the LC group and CT. (B) Correlation between seven core genes and immune cells. (C,~D) NMF analysis. (E) Heat map of genetic differences in the two subtypes. (F) Differential expression of LC-Endothelial gene in two subgroups.}
  \label{fig:fig7}
\end{figure}

To further characterise the disease relevance of LC-Endothelial--associated gene signatures, comparative expression and clustering analyses were performed. The expression levels of the 25 LC-Endothelial core genes were first compared between cirrhotic (LC) and control (CT) groups (Figure~\ref{fig:fig7}A). The results demonstrated a significant upregulation of these genes in the cirrhotic group compared with controls ($P = 6.3 \times 10^{-11}$), indicating a strong association between LC-Endothelial gene expression patterns and liver cirrhosis status. This differential expression supports the consistency of the LC-Endothelial signature identified in previous analyses.

To further investigate the relationship between LC-Endothelial gene expression and the cellular microenvironment, a heatmap was generated to visualise immune and stromal cell infiltration patterns associated with differential gene expression (Figure~\ref{fig:fig7}B). Increased enrichment of fibroblasts, endothelial cells, cytotoxic lymphocytes, T cells, myeloid dendritic cells, and natural killer (NK) cells was observed in association with elevated expression of key LC-Endothelial genes, including HSPB1, GADD45A, ATP1B3, ENPP2, and PARL. These patterns suggest coordinated changes between endothelial-associated gene activity and immune--stromal cell composition in cirrhotic tissues.

To explore molecular heterogeneity within the cirrhotic patient cohort, non-negative matrix factorisation (NMF) was applied to gene expression data to identify latent molecular subtypes. Patients were robustly partitioned into two distinct molecular subtypes (Subtype~1 and Subtype~2). Subtype~2 was characterised by higher expression of LC-Endothelial--associated genes, whereas Subtype~1 exhibited comparatively lower expression levels. The robustness of this clustering was supported by the consensus matrix and a silhouette score of 0.86, indicating well-separated and stable subtypes (Figure~\ref{fig:fig7}C,~D).

The gene-wise contribution to each subtype was assessed based on NMF coefficients and sparseness measures. Distinct gene contribution patterns were observed between subtypes: genes such as HSPB1 and FADS3 contributed more prominently to Subtype~1, whereas PSMB5 and MGST2 showed higher contributions in Subtype~2 (Figure~\ref{fig:fig7}E). These differences highlight intra-disease heterogeneity and suggest that LC-Endothelial--associated gene expression captures meaningful variation in molecular states among cirrhotic patients.

Finally, differential expression analysis of the 25 LC-Endothelial core genes between Subtype~1 and Subtype~2 further confirmed subtype-specific patterns (Figure~\ref{fig:fig7}F). LC-Endothelial gene expression was significantly higher in Subtype~2 than in Subtype~1 ($P = 0.012$), reinforcing the association between LC-Endothelial signatures and molecular stratification of cirrhosis. Together, these results demonstrate that LC-Endothelial--related gene expression patterns are consistently linked to immune--stromal composition and molecular subtypes, supporting their utility for downstream predictive modelling and disease stratification.

\subsection{Molecular docking and prediction of potential drugs}
Beyond disease diagnosis, potential therapeutic relevance of the identified LC-Endothelial--associated genes was explored through molecular docking analysis as a complementary in silico validation step. Owing to the limited availability of experimentally resolved protein structures and corresponding drug candidates, C1QBP and ENPP2 were selected for downstream docking analysis based on their biological relevance and database accessibility.

\begin{figure}[!htbp]
  \centering
  \includegraphics[width=\linewidth,height=0.78\textheight,keepaspectratio]{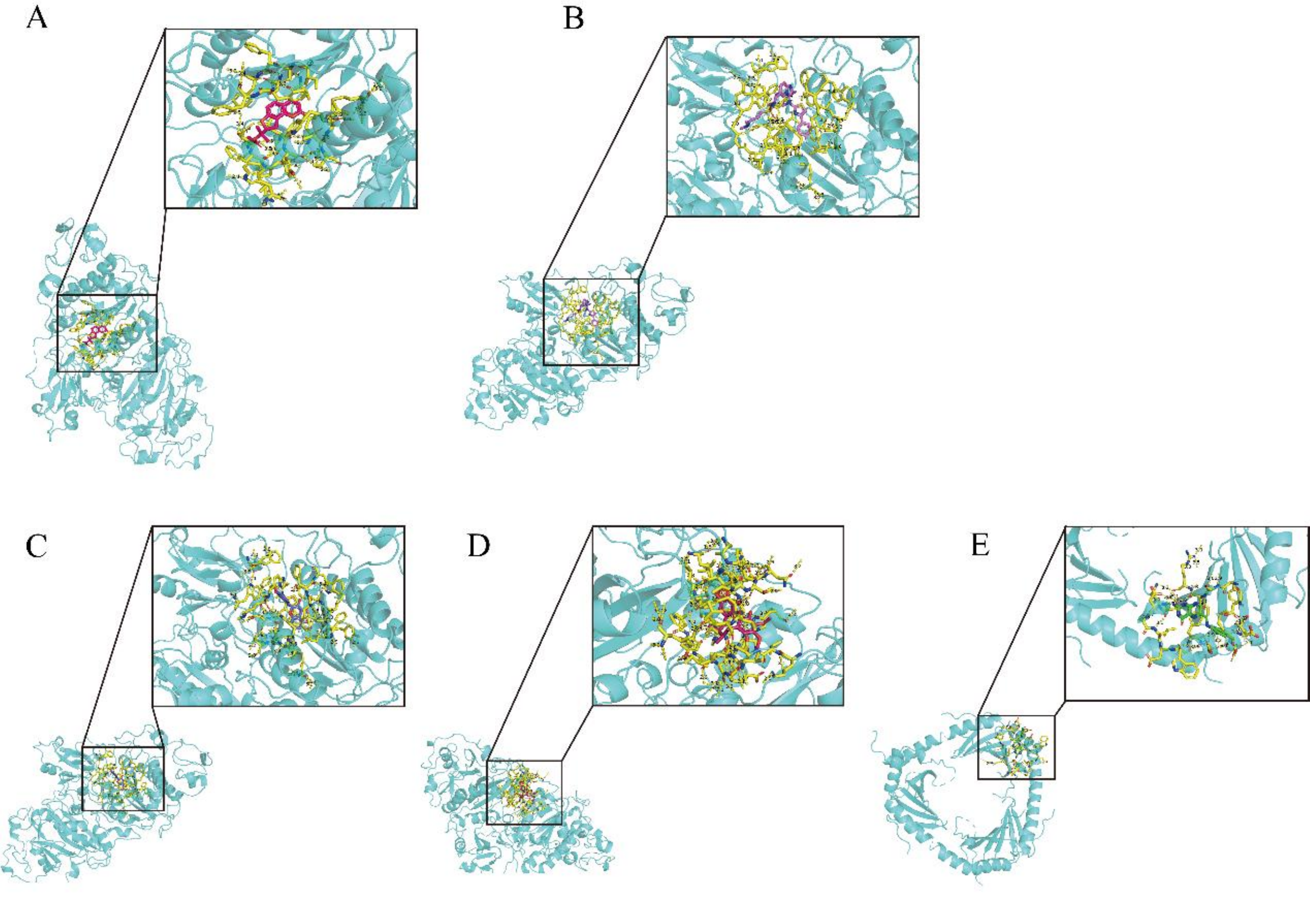}
  \caption{The screening of drugs. (A--D) ENPP2 protein with: (A) Schizandrin~B, (B) Fenofibrate, (C) Genistein, (D) Quercetin. (E) C1QBP protein with Folic Acid. The blue section represents the protein structure; the solid yellow section represents the binding site around the small molecule; the dashed yellow section is the hydrogen bond and the number is the strength of the hydrogen bond; the hydrogen bonds connect the drug to the amino acid residues on the protein, and the small molecule in the middle of the yellow region is the acting drug.}
  \label{fig:fig8}
\end{figure}

ENPP2 has been reported to play a facilitating role in liver cirrhosis progression~\cite{ref29}. To evaluate the feasibility of pharmacological modulation of ENPP2, candidate small-molecule compounds were screened by simulating docking interactions between ENPP2 macromolecules and drug ligands (Figure~\ref{fig:fig8}). The docking results demonstrated stable binding conformations with favourable binding energies, suggesting potential molecular compatibility between ENPP2 and the selected compounds.

C1QBP is considered a contributing factor in cirrhosis, particularly through its involvement in hepatic stellate cell (HSC) activation~\cite{ref30}. Using the same docking strategy, a candidate compound targeting C1QBP was identified. The predicted binding mode exhibited a stable interaction between the ligand and the C1QBP protein (Figure~\ref{fig:fig8}E), indicating a feasible binding relationship at the molecular level.

The calculated binding energies for the docked complexes were $-6.691$ (Figure~\ref{fig:fig8}A), $-9.969$ (Figure~\ref{fig:fig8}B), $-8.871$ (Figure~\ref{fig:fig8}C), $-8.592$ (Figure~\ref{fig:fig8}D), and $-6.18$ (Figure~\ref{fig:fig8}E). These values fall within an energetically favourable range, supporting the structural plausibility of ligand--protein interactions. It should be noted that molecular docking provides a predictive assessment of binding feasibility rather than direct evidence of pharmacodynamic or therapeutic efficacy.

Overall, this docking analysis serves as a proof-of-concept extension of the ML framework, demonstrating that key LC-Endothelial--associated genes identified through integrative analysis may also represent structurally druggable targets. These findings provide a computational basis for future experimental validation and therapeutic exploration in liver cirrhosis.

%================== Discussion ==================
\section{Discussion}\label{sec:discussion}

Liver cirrhosis represents a prototypical complex disease characterised by heterogeneous aetiology, progressive cellular remodelling, and limited opportunities for early intervention. Although conventional diagnostic strategies based on laboratory indicators, imaging, and biopsy remain clinically important, their effectiveness is constrained by invasiveness, sampling bias, and limited sensitivity at early disease stages~\cite{ref2,ref31}. From a computational perspective, cirrhosis poses substantial challenges due to high-dimensional molecular measurements, strong feature correlations, biological noise, and limited labelled samples~\cite{ref3}. These characteristics make it an appropriate testbed for evaluating ML strategies designed to operate under uncertainty and data heterogeneity.

Using single-cell transcriptomic analysis, this study identified endothelial cells as a disease-associated cellular population in liver cirrhosis~\cite{ref10}. Beyond their established biological relevance in vascular remodelling and fibrogenesis, endothelial cells offer a computationally advantageous entry point by capturing coordinated molecular and microenvironmental changes associated with disease progression. Alterations in endothelial function---such as sinusoidal capillarisation, dysregulated angiogenesis, and enhanced inflammatory signalling---are reflected in both gene expression patterns and intercellular communication networks~\cite{ref32,ref33,ref34}. These properties make endothelial cells a stable and informative substrate for downstream network-based feature extraction and modelling.

To characterise endothelial cell involvement at the molecular level, seven LC-Endothelial--associated core genes (HSPB1, GADD45A, CLDN5, ATP1B3, C1QBP, ENPP2, and PARL) were identified through a combination of network-based stabilisation and sparse feature selection. Collectively, these genes capture key aspects of cellular stress response, endothelial barrier integrity, immune regulation, metabolic dysfunction, and mitochondrial homeostasis~\cite{ref35,ref36}. Their biological relevance to fibrosis and inflammation has been supported by previous studies, reinforcing the validity of the network-driven feature extraction strategy.

Importantly, pathway-level analyses further demonstrated that these molecular signatures are embedded within broader regulatory processes, including CCL-mediated chemokine signalling and membrane lipid metabolic pathways. From a systems perspective, these pathways highlight how endothelial cells integrate inflammatory recruitment, metabolic remodelling, and fibrotic activation through interconnected genetic and signalling mechanisms~\cite{ref37,ref38}. Rather than serving solely as biological interpretation, these findings support the coherence and stability of the selected feature space for computational modelling.

To translate stabilised molecular signatures into diagnostic decision-making, this study combined sparse learning with both classical ML and representation-enhanced deep learning~\cite{ref39}. The seven LC-Endothelial feature genes demonstrated consistent classification performance across multiple ML models, indicating robustness to model-specific inductive biases.

Building upon this, a CNN--based disease map strategy was introduced to capture non-linear interactions between molecular signatures and immune microenvironment features~\cite{ref17}. Unlike conventional deep learning approaches applied directly to high-dimensional omics data, the CNN in this framework operates on structured two-dimensional representations constructed from a compact, biologically constrained feature set~\cite{ref16}. By integrating only sparsely selected genes with immune infiltration profiles, the effective model capacity is substantially reduced, mitigating over-parameterisation and lowering the risk of overfitting.

From a ML perspective, the CNN component functions primarily as a representation aggregation mechanism rather than an unconstrained black-box predictor. Its role is to model approximate, non-linear relationships under uncertainty while maintaining interpretability at the feature level. The use of independent training and validation cohorts further supports the generalisability of the learned representations and reduces the likelihood that the observed performance reflects dataset-specific artefacts or information leakage.

From a translational perspective, the identified LC-Endothelial signature genes~\cite{ref40} provide a feasible entry point for clinical deployment. The seven core genes represent a compact and interpretable molecular panel that is amenable to routine laboratory assays. In particular, these genes could be translated into diagnostic biomarkers through quantitative PCR or targeted transcriptomic panels using liver biopsy specimens or circulating nucleic acids, enabling cost-effective and scalable assessment in clinical settings. Rather than serving as standalone diagnostic tests, such molecular signatures could function as adjunctive tools to support early risk stratification and disease monitoring alongside existing laboratory and imaging-based assessments.

In addition, the molecular docking analysis performed in this study provides a preliminary decision-support layer for therapeutic exploration. The predicted interactions between selected candidate compounds and key molecular targets suggest structural feasibility for pharmacological modulation. Importantly, these results do not imply immediate clinical applicability but may inform hypothesis-driven drug repositioning or prioritisation strategies. In practice, such computational insights could be integrated with existing treatment paradigms---such as antifibrotic therapies, anti-inflammatory agents, or interventions targeting portal hypertension---to guide preclinical validation and experimental design rather than replacing established clinical protocols.

Beyond transcriptomic data, the proposed framework is inherently extensible to other biomedical modalities, highlighting its potential for multi-modal soft computing applications. For example, imaging-derived features from CT or MRI, including radiomic descriptors of liver morphology, texture, and vascular structure, could be incorporated as additional representation channels within the framework. By integrating molecular, immune, and imaging features into unified structured representations, the framework could capture complementary disease characteristics across multiple abstraction levels. Such multi-modal integration may further enhance diagnostic robustness and reduce modality-specific uncertainty, particularly in heterogeneous diseases such as liver cirrhosis.

Overall, the proposed framework illustrates how soft computing principles can support the translation of complex biomedical data into clinically meaningful decision-support tools. By combining modular design, tolerance to uncertainty, and multi-modal extensibility, the framework provides a flexible foundation for future studies aimed at bridging computational disease modelling and real-world clinical applications.

Despite the encouraging diagnostic performance observed, several limitations should be acknowledged. As with many transcriptomic studies, sample size remains limited, which constrains the statistical power available for deep learning--based approaches. Although the representation-enhanced design of the CNN reduces model complexity, future validation using larger, multi-centre datasets will be essential to further assess robustness and generalisability across diverse clinical settings.

%================== Conclusion ==================
\section{Conclusion}\label{sec:conclusion}

This study presents a unified ML framework that integrates single-cell analysis, network-based feature stabilisation, sparse learning, representation-enhanced deep learning, and decision-support modelling to address the challenges of complex disease diagnosis. By identifying a functionally distinct LC-Endothelial subpopulation and its associated molecular signature, the proposed framework provides a computationally robust approach for modelling liver cirrhosis under uncertainty. More broadly, this work demonstrates how ML principles can be applied to integrate heterogeneous biomedical data and support both diagnostic and translational exploration in complex disease settings.

%================== Back matter ==================
\section*{CRediT authorship contribution statement}
\textbf{Xueyuan Huang:} Writing -- review \& editing, Writing -- original draft, Supervision, Conceptualization.
\textbf{Yuheng Wang:} Writing -- original draft, Software, Methodology, Investigation, Conceptualization.
\textbf{Yuanzhi He:} Writing -- original draft, Visualization, Software, Formal analysis.
\textbf{Siqi Gou:} Writing -- review \& editing, Validation, Supervision, Formal analysis.
\textbf{Lu Bai:} Visualization, Validation.
\textbf{Wenqian Wu:} Visualization, Validation.
\textbf{Peifeng Liu:} Writing -- review \& editing.
\textbf{Aijia Wang:} Data curation.
\textbf{Tianhui Fan:} Resources, Project administration.
\textbf{Ze Zhou:} Investigation, Validation.
\textbf{Jiayu Xu:} Visualization, Validation, Resources, Project administration, Data curation, Writing -- review \& editing.

\section*{Acknowledgments}
Not applicable.

\section*{Data availability statement}
The datasets analysed in the current study are available in the GEO repository (\url{https://www.ncbi.nlm.nih.gov/geo/}, accessed on 10 August 2024).

%================== Bibliography ==================
\bibliographystyle{unsrtnat}
\bibliography{main}

\end{document}